Schrödinger equation for classical particles

M. Kozlowski[1,*], J. Marciak – Kozlowska[2]

Magdalena Pelc[3]

[1] Warsaw University,  [2] Institute of Electron Technology

[3] Elementary School, Raclawicka 14, Warsaw, Poland

Abstract

In this paper we propose the hyperbolic Schrödinger equation (HS). The solution of the HS for a particle in a box is obtained. It is shown that for particles with $m >> M_p$ the energy spectrum is independent of the mass of particle.

[*] corresponding author: e-mail: miroslawkozlowski@aster.pl

1. Modified Schrödinger Equation

When M. Planck made the first quantum discovery he noted an interesting fact [1]. The speed of light, Newton's gravity constant and Planck's constant clearly reflect fundamental properties of the world. From them it is possible to derive the characteristic mass $M_P$, length $L_P$ and time $T_P$ with approximate values

$L_P = 10^{-35}$ m

$T_P = 10^{-43}$ s

$M_P = 10^{-5}$ g .

Nowadays much of cosmology is concerned with "interface" of gravity and quantum mechanics.

In this paper we investigate the question: how gravity can modify the quantum mechanics, i.e. the nonrelativistic Schrödinger equation (SE). We argue that SE with relaxation term describes properly the quantum behavior of particle with mass $m_i < M_P$ and contains the part which can be interpreted as the pilot wave equation. For $m_i \quad M_P$ the solution of the SE represent the strings with mass $M_P$.

The thermal history of the system (heated gas container, semiconductor or Universe) can be described by the generalized Fourier equation [2]

$$q(t) = -\int_{-\infty}^{t} \underbrace{K(t-t')}_{\text{thermal history}} \underbrace{\nabla T(t') dt'}_{\text{diffusion}}.$$

(1)

In Eq. (1) $q(t)$ is the density of the energy flux, $T$ is the temperature of the system and $K(t - t')$ is the thermal memory of the system

$$K(t-t') = \frac{K}{\tau} \exp\left[-\frac{(t-t')}{\tau}\right],$$

(2)

where $K$ is constant, and $\tau$ denotes the relaxation time.

As was shown in [2]

$$K(t-t') = \begin{cases} K\delta(t-t') & \text{diffusion} \\ K = \text{constant} & \text{wave} \\ \dfrac{K}{\tau}\exp\left[-\dfrac{(t-t')}{\tau}\right] & \text{damped wave} \\ & \text{or hyperbolic diffusion.} \end{cases}$$

The damped wave or hyperbolic diffusion equation can be written as:

$$\frac{\partial^2 T}{\partial t^2} + \frac{1}{\tau}\frac{\partial T}{\partial t} = \frac{D_T}{\tau}\nabla^2 T.$$

(3)

For $\tau \to 0$, Eq. (3) is the Fourier thermal equation

$$\frac{\partial T}{\partial t} = D_T \nabla^2 T$$

(4)

and $D_T$ is the thermal diffusion coefficient. The systems with very short relaxation time have very short memory. On the other hand for $\tau \to \infty$ Eq. (3) has the form of the thermal wave (undamped) equation, or *ballistic* thermal equation. In the solid state physics the *ballistic* phonons or electrons are those for which $\tau \to \infty$. The experiments with *ballistic* phonons or electrons demonstrate the existence of the *wave motion* on the lattice scale or on the electron gas scale.

$$\frac{\partial^2 T}{\partial t^2} = \frac{D_T}{\tau}\nabla^2 T.$$

(5)

For the systems with very long memory Eq. (3) is time symmetric equation with no arrow of time, for the Eq. (5) does not change the shape when $t \to -t$.

In Eq. (3) we define:

$$= \left(\frac{D_T}{\ }\right),$$

(6)

velocity of thermal wave propagation and

$$= ,$$

(7)

where is the mean free path of the heat carriers. With formula (6) equation (3) can be written as

$$\frac{1}{\ ^2}\frac{\partial^2 T}{\partial t^2} + \frac{1}{\ ^2}\frac{\partial T}{\partial t} = \nabla^2 T.$$

(8)

From the mathematical point of view equation:

$$\frac{1}{\ ^2}\frac{\partial^2 T}{\partial t^2} + \frac{1}{D}\frac{\partial T}{\partial t} = \nabla^2 T$$

is the hyperbolic partial differential equation (PDE). On the other hand Fourier equation

$$\frac{1}{D}\frac{\partial T}{\partial t} = \nabla^2 T$$

(9)

and Schrödinger equation

$$i\hbar\frac{\partial \Psi}{\partial t} = -\frac{\hbar^2}{2m_i}\nabla^2 \Psi$$

(10)

are the parabolic equations. Formally with substitutions

$$t \leftrightarrow it, \ \Psi \leftrightarrow T.$$

(11)

Fourier equation (9) can be written as

$$i\hbar \frac{\partial \Psi}{\partial t} = -D\hbar \nabla^2 \Psi$$

(12)

and by comparison with Schrödinger equation one obtains

$$D_T \hbar = \frac{\hbar^2}{2m_i}$$

(13)

and

$$D_T = \frac{\hbar}{2m_i}.$$

(14)

Considering that $D_T = \upsilon^2 \tau$ (6) we obtain from (14)

$$\tau = \frac{\hbar}{2m_i \upsilon_h^2}.$$

(15)

Formula (15) describes the relaxation time for quantum thermal processes.

Starting with Schrödinger equation for particle with mass $m_i$ in potential $V$:

$$i\hbar \frac{\partial \Psi}{\partial t} = -\frac{\hbar^2}{2m_i} \nabla^2 \Psi + V\Psi$$

(16)

and performing the substitution (11) one obtains

$$\hbar \frac{\partial T}{\partial t} = \frac{\hbar^2}{2m_i} \nabla^2 T - VT$$

(17)

$$\frac{\partial T}{\partial t} = \frac{\hbar}{2m_i}\nabla^2 T - \frac{V}{\hbar}T.$$
(18)

Equation (18) is Fourier equation (parabolic PDE) for $\tau = 0$. For $\tau \neq 0$ we obtain

$$\tau\frac{\partial^2 T}{\partial t^2} + \frac{\partial T}{\partial t} + \frac{V}{\hbar}T = \frac{\hbar}{2m_i}\nabla^2 T,$$
(19)

$$\tau = \frac{\hbar}{2m_i\upsilon^2}$$
(20)

or

$$\frac{1}{\upsilon^2}\frac{\partial^2 T}{\partial t^2} + \frac{2m_i}{\hbar}\frac{\partial T}{\partial t} + \frac{2Vm_i}{\hbar^2}T = \nabla^2 T.$$

With the substitution (11) equation (19) can be written as

$$i\hbar\frac{\partial \Psi}{\partial t} = V\Psi - \frac{\hbar^2}{2m_i}\nabla^2\Psi - \tau\hbar\frac{\partial^2 \Psi}{\partial t^2}.$$
(21)

The new term, relaxation term

$$\tau\hbar\frac{\partial^2 \Psi}{\partial t^2}$$
(22)

describes the interaction of the particle with mass $m_i$ with space-time. When the quantum particle is moving through the quantum void it is influenced by scaterring on the virtual electron=-positron pairs The relaxation time $\tau$ can be calculated as:

$$\tau^{-1} = \left(\tau_{e-p}^{-1} + ... + \tau_{Planck}^{-1}\right),$$
(23)

where, for example $\tau_{e\text{-}p}$ denotes the scattering of the particle $m_i$ on the electron-positron pair ($\tau_{e-p} \sim 10^{-17}$ s) and the shortest relaxation time $\tau_{Planck}$ is the Planck time ($\tau_{Planck} \sim 10^{-43}$ s).

From equation (23) we conclude that $\tau \approx \tau_{Planck}$ and equation (21) can be written as

$$i\hbar \frac{\partial \Psi}{\partial t} = V\Psi - \frac{\hbar^2}{2m_i}\nabla^2\Psi - \tau_{Planck}\hbar\frac{\partial^2\Psi}{\partial t^2},$$

(24)

where

$$\tau_{Planck} = \frac{1}{2}\left(\frac{\hbar G}{c^5}\right)^{\frac{1}{2}} = \frac{\hbar}{2M_p c^2}.$$

(25)

In formula (25) $M_p$ is the mass Planck. Considering Eq. (25), Eq. (24) can be written as

$$i\hbar\frac{\partial \Psi}{\partial t} = -\frac{\hbar^2}{2m_i}\nabla^2\Psi + V\Psi - \frac{\hbar^2}{2M_p}\nabla^2\Psi$$
$$+ \frac{\hbar^2}{2M_p}\nabla^2\Psi - \frac{\hbar^2}{2M_p c^2}\frac{\partial^2\Psi}{\partial t^2}.$$

(26)

The last two terms in Eq. (26) can be defined as the *Bohmian* pilot wave

$$\frac{\hbar^2}{2M_p}\nabla^2\Psi - \frac{\hbar^2}{2M_p c^2}\frac{\partial^2\Psi}{\partial t^2} = 0,$$

(27)

i.e.

$$\nabla^2\Psi - \frac{1}{c^2}\frac{\partial^2\Psi}{\partial t^2} = 0.$$

(28)

It is interesting to observe that pilot wave $\Psi$ does not depend on the mass of the particle. With postulate (28) we obtain from equation (26)

$$i\hbar \frac{\partial \Psi}{\partial t} = -\frac{\hbar^2}{2m_i}\nabla^2\Psi + V\Psi - \frac{\hbar^2}{2M_p}\nabla^2\Psi$$

(29)

and simultaneously

$$\frac{\hbar^2}{2M_p}\nabla^2\Psi - \frac{\hbar^2}{2M_p c^2}\frac{\partial^2\Psi}{\partial t^2} = 0.$$

(30)

In the operator form Eq. (21) can be written as

$$\hat{E} = \frac{\hat{p}^2}{2m_i} + \frac{1}{2M_p c^2}\hat{E}^2,$$

(31)

where $\hat{E}$ and $\hat{p}$ denote the operators for energy and momentum of the particle with mass $m_i$. Equation (31) is the new dispersion relation for quantum particle with mass $m_i$. From Eq. (21) one can concludes that Schrödinger quantum mechanics is valid for particles with mass $m_i \ll M_P$. But pilot wave exists independent of the mass of the particles.

For particles with mass $m_i \ll M_P$ Eq. (29) has the form

$$i\hbar \frac{\partial \Psi}{\partial t} = -\frac{\hbar^2}{2m_i}\nabla^2\Psi + V\Psi.$$

(32)

In the case when $m_i \approx M_p$ Eq. (29) can be written as

$$i\hbar \frac{\partial \Psi}{\partial t} = -\frac{\hbar^2}{2M_p}\nabla^2\Psi + V\Psi,$$

(33)

but considering Eq. (30) one obtains

$$i\hbar \frac{\partial \Psi}{\partial t} = -\frac{\hbar^2}{2M_p c^2} \frac{\partial^2 \Psi}{\partial t^2} + V\Psi$$

(34)

or

$$\frac{\hbar^2}{2M_p c^2} \frac{\partial^2 \Psi}{\partial t^2} + i\hbar \frac{\partial \Psi}{\partial t} - V\Psi = 0.$$

(35)

2. Gravity and Schrödinger Equation

Classically, when the inertial mass $m_i$ and the gravitational mass $m_g$ are equated the mass drops out of Newton's equation of motion, implying that particles of different mass with the same initial condition follows the same trajectories. But in Schrödinger's equation the masses do not cancel. For example in a uniform gravitational field [2]

$$i\hbar \frac{\partial \Psi}{\partial t} = -\frac{\hbar^2}{2m_i} \frac{\partial^2 \Psi}{\partial x^2} + m_g g x \Psi$$

(36)

implying mass dependent difference in motion.

In this paragraph we investigate the motion of particle with inertial mass $m_i$ in the potential field $V$. The potential $V$ contains all the possible interactions including the gravity.

$$i\hbar \frac{\partial \Psi}{\partial t} = V\Psi - \frac{\hbar^2}{2m_i} \nabla^2 \Psi - 2\tau\hbar \frac{\partial^2 \Psi}{\partial t^2}$$

(37)

where the term

$$2\tau\hbar\frac{\partial^2 \Psi}{\partial t^2}, \quad \tau = \frac{\hbar}{m_i c^2}$$

(38)

describes the memory of the particle with mass $m_i$. Above equation for the wave function is the local equation with finite invariant speed, $c$ which equals the light speed in the vacuum.

Let us look for the solution of the Eq. (26), $V=0$, in the form (for 1D)

$$\Psi = \Psi(x - ct).$$

(39)

For 0, i.e. for finite Planck mass we obtain:

$$\Psi(x - ct) = \exp\left(\frac{2\mu ic}{\hbar}(x - ct)\right)$$

(40)

where the reduced $\mu$ mass equals

$$\mu = \frac{m_i M_p}{m_i + M_p}$$

(41)

For $m_i \ll M_p$, i.e. for all elementary particles one obtains

$$\mu = m_i \tag{42}$$

and formula (40) describes the wave function for free Schrödinger particles

$$\Psi(x - ct) = \exp\left(\frac{2m_i ic}{\hbar}(x - ct)\right)$$

(43)

For $m_i \gg M_p$, $\mu = M_p$

$$\Psi(x-ct) = \exp\left(\frac{2M_p ic}{\hbar}(x-ct)\right)$$

(44)

From formula (44) we conclude that $\Psi(x-ct)$ is independent of mass $m_i$. In the case $m_i < M_p$ from formulae (41) and (42) one obtains

$$\mu = m_i\left(1 - \frac{m_i}{M_p}\right)$$

$$\Psi(x-ct) = \exp\left(\frac{2im_i c}{\hbar}(x-ct)\right)\exp\left(-i\frac{m_i}{M_p}\left(\frac{2m_i c}{\hbar}x - \frac{2m_i c^2}{\hbar}t\right)\right)$$

(45)

In formula (45) we put

$$k = \frac{2m_i c}{\hbar}$$

$$\omega = \frac{2m_i c^2}{\hbar}$$

(46)

and obtain

$$\Psi(x-ct) = e^{i(kx-\omega t)} e^{-i\frac{m_i}{M_p}(kx-\omega t)}$$

(47)

As can concluded from formula (47) the second term depends on the gravity

$$\exp\left[-i\frac{m_i}{M_p}(kx-\omega t)\right] = \exp\left[-i\left(\frac{m_i^2 G}{\hbar c}\right)^{1/2}(kx-\omega t)\right]$$

(48)

where $G$ is the Newton gravity constant.

It is interesting to observe that the new constant, $\alpha_G$,

$$\alpha_G = \frac{m_i^2 G}{\hbar c} \quad (49)$$

is the gravitational constant. For $m_i = m_N$ nucleon mass

$$\alpha_G = 5.9042 \cdot 10^{-39} \quad (50)$$

3. The Particle in a Box

This quantum mechanical system by a particle of mass $m$ confined in a one-dimensional box of length $L$. For the particle to be confined within region II, the potential energy outside (regions I and III) is assumed to be infinite. In order to understand further this system, we need to formulate and solve the Schrödinger equations (26).

$$i\hbar \frac{\partial \Psi}{\partial t} = -\frac{\hbar^2}{2m_i}\nabla^2\Psi + V\Psi - \frac{\hbar^2}{2M_p}\nabla^2\Psi + \frac{\hbar^2}{2M_p}\left(\nabla^2\Psi - \frac{1}{c^2}\frac{\partial^2\Psi}{\partial t^2}\right). \quad (51)$$

Considering the pilot wave equation

$$\nabla^2\Psi - \frac{1}{c^2}\frac{\partial^2\Psi}{\partial t^2} = 0 \quad (52)$$

one obtains

$$i\hbar \frac{\partial \Psi}{\partial t} = -\frac{\hbar^2}{2\mu}\nabla^2\Psi + V\Psi, \quad (53)$$

where

$$\mu = \frac{m_i M_p}{m_i + M_p}.$$

| In region II, $V(x) = 0$ | In regions I and III, $V(x) = \infty$ |
|---|---|
| $-\dfrac{\hbar^2}{2\mu}\nabla^2\Psi + V\Psi = E\Psi$ | $-\dfrac{\hbar^2}{2\mu}\nabla^2\Psi + V\Psi = E\Psi$ |
| $-\dfrac{\hbar^2}{2\mu}\nabla^2\Psi + 0 = E\Psi$ | $-\dfrac{\hbar^2}{2\mu}\nabla^2\Psi + \infty\Psi = E\Psi$ |
| $-\dfrac{\hbar^2}{2\mu}\nabla^2\Psi = E\Psi$ | $-\dfrac{\hbar^2}{2\mu}\nabla^2\Psi = (E - \infty)\Psi$ |

(54)

For regions I and III (outside the box), the solution is straightforward, the wavefunction $\Psi$ is zero.

For region II (inside the box), we need to find a function that regenerates itself after taking its second derivative.

$$-\frac{\hbar^2}{2\mu}\nabla^2\Psi = E\Psi$$

$$\nabla^2\Psi = \frac{2\mu E}{\hbar^2}\Psi = -k^2\Psi, \quad \text{where we define } k = \sqrt{\frac{2\mu E}{\hbar^2}}.$$

(55)

Perfect candidates would be the trigonometric sine and cosine functions.

$$\Psi(x) = A\cos(kx) + B\sin(kx).$$

(56)

To further refine the wavefunction, we need to impose boundary conditions: At $x=0$, the wavefunction should be zero.

$$\Psi(0) = A\cos(k0) + B\sin(k0) = A \cdot 1 + B \cdot 0.$$

(57)

Equation can only be true if $A = 0$: $\Psi(x) = B\sin(kx).$

In addition, at $x=L$, the wavefunction should also be zero.

$$\Psi(0) = 0 = B\sin(kL). \quad \text{True when } kL = n\pi \text{ or } k = \frac{n\pi}{L}:$$

$$\Psi(x) = B\sin(\frac{n\pi}{L}x), \quad n = 1, 2, 3, ...$$

(58)

Going back to the Schrödinger's equation, we can then formulate the energies

$$kL = n\pi \quad \text{and} \quad k = \sqrt{\frac{2\mu E}{\hbar^2}}$$

$$\sqrt{\frac{2\mu E}{\hbar^2}} = \frac{n\pi}{L} \implies \frac{2\mu E}{\hbar^2} = \frac{n^2\pi^2}{L^2}$$

$$E_n = \frac{n^2\pi^2\hbar^2}{2\mu L^2} = \frac{n^2 h^2}{8\mu L^2} \quad \text{since } \hbar = \frac{h}{2\pi}.$$

(59)

Thus, the application of the Schrödinger equation to this problem results in the well known expressions for the wavefunctions and energies, namely:

$$\Psi_n = \sqrt{\frac{2}{L}}\sin\left(\frac{n\pi x}{L}\right) \text{ and } E_n = \frac{n^2 h^2}{8\mu L^2}.$$

(60)

From formula (60) we conclude that for "heavy" classical particles, i.e. for $m_i >> M_p$ energy spectrum of the particle in the box is independent of the mass of particle

$$E_n = \frac{n^2 h^2}{8 M_p L^2}.$$

(61)

4. Conclusions

The hyperbolic Schrödinger equation when applied to the study of particle in a box offers new picture of the classical extension of the quantum mechanics. The classical particles, i.e. particles with $m_i >> M_p$ have distinct energy spectrum which is independent of its mass.